\documentclass[aps,prl,twocolumn,showpacs, preprintnumbers]{revtex4}
\usepackage{dcolumn}
\usepackage{bm}
\usepackage{epsfig}
\usepackage{citesort}

\newcommand{\etal}{{\it et al.}}

\begin{document}

\preprint{\tighten\vbox{\hbox{\hfil SUHEP-06-2006}
                       \hbox{\hfil Aug., 2009}}}

\title{Model Independent Methods for
Determining ${\cal{B}}\left(\Upsilon(5S) \to
B_S^{(*)}\overline{B}_S^{(*)}\right)$}

\author{Radia~Sia}
\author{Sheldon~Stone}
\affiliation {Department of Physics, Syracuse University,
Syracuse, New York 13244-1130}

\date{August 12, 2009}

\begin{abstract}
We describe a method that provides a model independent measurement
of the $B_S$ fraction in $\Upsilon$(5S) resonance decays, $f_S$,
using the relative rates of like-sign versus opposite sign
dileptons; the like-sign leptons result from $B^0$ and $B_S$
mixing. In addition, we show that determining the rates of single,
double and triple $D_S^{\pm}$ mesons provides an alternative way
of finding $f_S$.

\end{abstract}
 \maketitle

The $\Upsilon$(5S) resonance has long been thought to be a source
of $B_S$ mesons, since it is massive enough to produce
$B_S\overline{B}_S$ pairs \cite{Ono}. Recently, the CLEO
collaboration established the presence of $B_S$ mesons and made a
model dependent measurement that the $B_S$ fraction, $f_S$, of
$\Upsilon(5S)$ decays is (16.0$\pm$2.6$\pm$5.8)\%, using a
theoretical estimate of (92$\pm$11)\% for the inclusive branching
ratio ${\cal{B}}(B_S\to D_S X)$ \cite{CLEO-DS}. Explicit $B_S$
final states have also been reconstructed \cite{CLEO-BS}. These
results have been confirmed by Belle \cite{Belle}.  Precision
measurements of $B_S$ branching fractions are one of the most
important goals of such studies. For example, it has been
suggested that measurements of ${\cal{B}}(B_S\to
D_S^{(*)+}D_S^{(*)-})$ lead to a determination of the lifetime
difference between CP+ and CP- eigenstates \cite{Drut}. To measure
these, however, it is imperative that the number of $B_S$ mesons
produced in $\Upsilon(5S)$ decays be well known. The purpose of
this paper is to describe techniques that can be used to produce
accurate, model independent measurements of $f_ S$. (Note that
we are not considering $\Upsilon$(5S) decays to modes that do not contain $B$ mesons.)

Both $B$ and $B_S$ mesons can result from $\Upsilon(5S)$ decays. The
possible final states are $B^{(*)-}B^{(*)+}$,
$B^{(*)0}\overline{B}^{(*)0}$, $B^{(*)}\overline{B}^{(*)}\pi$,
$B\overline{B}\pi\pi$, and $B_S^{(*)0}\overline{B}_S^{(*)0}$, since
there is not sufficient energy to produce an extra pion.

Our first method for determining $f_S$ requires the measurement of
like-sign versus opposite-sign dileptons. High momentum leptons
from $B$ (or $B_S$) decays reflect the flavor of the parent;
positive leptons result from $B$ decays, while negative leptons
arise from $\overline{B}$ decays. We will assume here that the
minimum lepton momentum requirement is large enough so that
contamination from the decay sequence $B\to DX$, $D\to Y \ell\nu$
is negligible, or suitable corrections can be applied
\cite{dilep}.

This technique relies on the fact that $B_S$ mixing oscillations are
very rapid compared to $B_d$. The mass difference for $B^0$ mesons
is  $\Delta m_d = 0.509\pm0.0005$ ps$^{-1}$, while for $B_S$ mesons,
$\Delta m_S$ is limited at 90\% confidence level to be $>$16.6
ps$^{-1}$ \cite{PDG,D0}, and recently measured by CDF to be
$17.31^{+0.33}_{-0.18}\pm 0.07$ ps$^{-1}$ \cite{CDF}. $B_S$ mixing,
when the meson decays semileptonically, produces a relatively large
number of like-sign dilepton events that allows a measurement of
$B_S$ production \cite{Schneider}. We must also account for $B_d$
mixing and thus need to know the composition of the $B\overline{B}$
states, since the level of mixing will depend on the Charge
Conjugation (C) state of the $B^0\overline{B}^0$ system.

Hadrons containing $b$-flavor are produced in pairs in current
experiments. If the $B$ and $\overline{B}$ are not in an
eigenstate of C, then the probability for a $B^0$ to decay as a
$\overline{B}^0$, integrated over time, is given by
\begin{equation} \label{eq:hads}
P(B^0\to \overline{B}^0)=\chi=\frac{x^2}{2(1+x^2)},
\end{equation}
where the mixing parameter $x=\Delta m/\Gamma$.

 The ratio $R$ of mixed events to unmixed events is given by
\begin{equation}
R={{N_{B^0B^0}+N_{\overline{B}^0\overline{B}^0}}
\over{N_{B^0\overline{B}^0}+N_{\overline{B}^0B^0}}}~.
\end{equation}

For incoherent states then $
N_{B^0B^0}+N_{\overline{B}^0\overline{B}^0}$ is equal to
$2\chi(1-\chi)$ and $ N_{B^0\overline{B}^0}+N_{\overline{B}^0B^0}$
is equal to $\chi^2+(1-\chi)^2$.

For $B_S$ mesons the mixing oscillations are so fast that R equals
1, regardless of the C parity state of the $B_S\overline{B}_S$
pair. For the $B^0$, however, $R$ does depend on the initial
$B^0\overline{B}^0$ state. If the initial pair of $B$ mesons is in
a C odd configuration then \cite{Bigi-Sanda}
\begin{equation}\label{eq:Rminus}
R_-={x^2\over{2+x^2}}~,
\end{equation}
where the mixing parameter $x_d=\Delta m_d/\Gamma$ and is well
measured as 0.775$\pm$0.008 \cite{HFAG}.

For C even configurations we have
\begin{equation}
 R_+={{3x^2+x^4}\over{2+x^2+x^4}}~.
\end{equation}\textbf{}

The states containing charged $B$ pairs do not contribute to mixing.
We expect that there are an equal number of such states as neutral
non-strange $B$'s, since the mass difference is small and Coulomb
corrections are small even at the $\Upsilon$(4S) \cite{marciano}.
The $B^0\overline{B^0}$ and $B^{*0}\overline{B}^{*0}$ are both
produced in an L equals one state; thus the $B^0\overline{B^0}$ pair
is in a negative C parity state. The $B^0\overline{B^0}\gamma$
state, (from $B^0\overline{B^0}^*$ or $B^{0*}\overline{B^0}$) on the other hand, is in a positive C parity state.

\begin{figure}
\centerline{\epsfig{figure=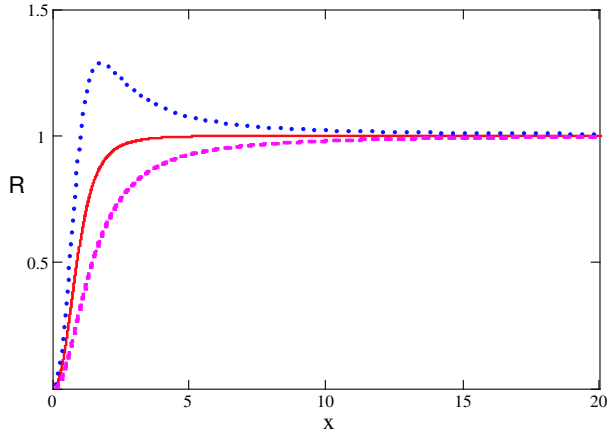,height=2.3in}} \vspace{-.2cm}
\caption{Ratio (R) of like-sign to opposite sign dileptons as a
function of $x=\Delta m/\Gamma$ for C odd (dashed), C even
(dotted) and incoherent (solid) $B^0\overline{B}^0$ or
$B_S\overline{B}_S$ states.}
\label{Rvsx}       
\end{figure}

Next we consider $B\overline{B}\pi$ and $B\overline{B}\pi\pi$ states.
The $B^0\overline{B^0}\pi\pi$ states can be incoherent, or coherent if
there is fixed angular momentum between the pions and between the
pions and the $B$'s .  $R$ for an
incoherent state is given in terms of $x$ as \cite{BGS}
\begin{equation}
 R_{\rm incho}=\frac{x^2(2+x^2)}{2+2x^2+x^4} ~.
\end{equation}
The variation of the $R$ functions with $x$ is
shown in Fig.~\ref{Rvsx}.
Note that $B^0B^-\pi^+$ and $\overline{B^0}B^+\pi^-$ final states contain only
one $B^0$ and mix according to Eq.~\ref{eq:hads}, leading to a mixing rate given
by Eq.~\ref{eq:Rminus}, and thus would be treated as an odd C state for our purpose.
The other states can also be classified in terms of C eigenstates or
incoherent states.

CLEO has made the first measurement of $B$ meson production at the
$\Upsilon$(5S) \cite{CLEO5S}. They find that
$B^{*0}\overline{B}^{*0}$ production is largest with
$B^{*0}\overline{B^0}$+$B^{0}\overline{B^{*0}}$ being about 1/3
its rate. Although limits on other decay channels are not small,
we will assume that more data will allow the exact composition of
the $B$ decays at the $\Upsilon$(5S) to be established.

\begin{figure}
\vspace{-.1cm} \centerline{\epsfig{figure=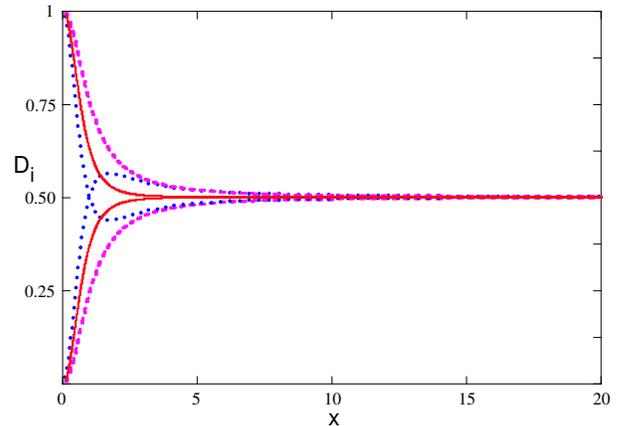,height=2.3in}}
\vspace{-.2cm} \caption{Functions for like-sign dileptons (rising
curves) and opposite-sign dileptons (falling curves) for C odd
(dashed), C even (dotted) and incoherent (solid)
$B^0\overline{B}^0$ or $B_S\overline{B}_S$ states.}
\label{Divsx}       
\end{figure}

We now consider like-sign dilepton production coming from neutral
$B$'s. The yield from $B^0\overline{B}^0$ pairs is given by
\begin{equation}
N_{++}+N_{--}=N_{5S}f_d{\cal{B}}_{d-sl}^2\sum_if_iD_{i-\pm\pm}(x),
\end{equation}
where $N_{5S}$ is the number of $\Upsilon(5S)$ events containing
a $B\overline{B}$ pair above
continuum background, and ${\cal{B}}_{d-sl}$ is the $B^0$
semileptonic branching ratio. The $D_{i-\pm\pm}(x)$ functions for
the neutral $B$ pairs in C=$\pm$ eigenstates, or being incoherent,
are listed in Table~\ref{functions}, and plotted in
Fig.~\ref{Divsx}. The fractions in each of these three states are
denoted by $f_i$; they include the $B\overline{B}\pi(\pi)$
final states \cite{special}. We also list and plot the $D_{i-\pm\mp}(x)$
functions for opposite sign dileptons. The fraction of neutral
non-strange $B$ mesons, $f_d$, is assumed to be equal to that for
charged $B$ mesons \cite{fdequalfc}. Since the sum of neutral,
charged, and strange $B$ mesons is unity, we have
\begin{eqnarray}\label{eq:fdfufs}
f_u+f_d+f_s&=&1\\\nonumber
2f_d+f_S&=&1, {\rm~and~therefore}\\\nonumber
f_d=f_u&=&(1-f_S)/2.
\end{eqnarray}

\begin{table}[htb]
\begin{center}
\begin{tabular}{lcc}
C state  &   $D_{i-\pm\pm}(x)$   &  $D_{i-\pm\mp}$ \\\hline
Odd & $\displaystyle\frac{x^2}{2(1+x^2)}$ & $\displaystyle\frac{(2+x^2)}{2(1+x^2)}$\\
& &\\
 Even & $\displaystyle\frac{x^2(3+x^2)}{2(1+x^2)^2}$ &
$\displaystyle\frac{(2+x^2+x^4)}{2(1+x^2)^2}$\\
&&\\
 Incoherent & $\displaystyle\frac{x^4+2x^2}{2(1+x^2)^2}$ &
$\displaystyle\frac{2+2x^2+x^4}{2(1+x^2)^2}$\\
\hline\hline
\end{tabular}
\end{center}
\caption{Functions for like-sign and opposite-sign dileptons.
\label{functions}}
\end{table}

 A similar set of expressions exist for
like-sign leptons from $B_S$ decays
\begin{equation}
N_{++}+N_{--}=N_{5S}f_S{\cal{B}}_{S-sl}^2\sum_if_iD_{i-\pm\pm}(x_s),
\end{equation}
where ${\cal{B}}_{S-sl}$ is the semileptonic branching ratio and
$D_{\pm\pm}(x_s)$ is the function that characterizes the dilepton
rate and, in principle, depends on whether or not
$B_S\overline{B}_S$ is in an even or odd eigenstate. The function
form is identical to the $D_{\pm\pm}(x)$ functions listed in
Table~\ref{functions}, but incoherent states are not allowed since
there isn't enough energy at the $\Upsilon(5S)$ to produce an
additional pion. Similarly, the opposite-sign functions for $B_S$,
$D_{\pm\mp}(x_s)$, are identical in form with the $D_{\pm\mp}(x)$
functions. Note that all these functions are normalized so that they
go to a value of 0.5 as $x$ (or $x_S$) gets large, reflecting the
fact that the mixing probability goes to its maximum value of 50\%.

The rate of opposite-sign dileptons is given by
\begin{eqnarray}
N_{+-}&+& N_{-+}=N_{5S}\left[
f_S{\cal{B}}_{S-sl}^2D_{\pm\mp}(x_s)\right. \\\nonumber
&&+\frac{1-f_S}{2}{\cal{B}}_{d-sl}^2\sum_i f_i D_{i-\pm\mp}
(x)+\frac{1-f_S}{2}\left.{\cal{B}}_{u-sl}^2 \right]~.
\end{eqnarray}

The last term is due to charged $B_u$ decays which do not mix. We
form the ratio of like-sign to opposite-sign dileptons and divide
through by the charged $B$ semileptonic branching ratio
${\cal{B}}_{u-sl}^2$. We replace the resulting  ratios of
semileptonic branching fractions by the lifetime
ratios, ${\cal{B}}_{d-sl}/{\cal{B}}_{u-sl}=\tau (B^0)/\tau
(B^+)=\tilde\tau_{+0}$, and
${\cal{B}}_{S-sl}/{\cal{B}}_{u-sl}=\tau (B_S)/\tau
(B^+)=\tilde\tau_{+S}$. Since the CDF measurement of $\Delta m_S$
when multiplied by the measured lifetime gives $x_S\approx$26.6
\cite{CDF}, we substitute $D_{\pm\pm}(x_S)=D_{\pm\mp}(x_S)$=0.5.
Denoting
\begin{equation}
\rho=\frac{N_{++}+N_{--}}{N_{+-}+N_{-+}}~,
\end{equation}
the resulting equation is
\begin{equation}
\rho=
\frac{f_S\tilde\tau_{+S}^2+(1-f_S)\tilde\tau_{+0}^2\sum_if_iD_{i-\pm\pm}(x)}
 {f_S\tilde\tau_{+S}^2+(1-f_S)\tilde\tau_{+0}^2\sum_if_iD_{i-\pm\mp}(x)
 +(1-f_S)}~.
\end{equation}

Solving for $f_S$ gives
\begin{equation}\displaystyle
f_S=\frac{S_D-\rho} {(\rho-1)\tilde\tau_{+S}^2+ S_D-\rho}~,
\end{equation}
where
\begin{equation}
S_D=\tilde\tau_{+0}^2\left[\sum_if_iD_{i-\pm\pm}(x)
-\rho\sum_if_iD_{i-\pm\mp}(x)\right]~.
\end{equation}

For practical application, we can use Eq.~\ref{eq:fdfufs} to
determine $f_d$ and $f_u$ and then use well measured $B^0$ and
$B^+$ branching ratios to normalize the $B_S$ rates.

We can estimate the error in $f_S$ by taking the C odd
contribution as 75\%, the C even component as 25\%, no incoherent
$B^0\overline{B}^0$ contribution and $f_S$ equal to 16\% from the
CLEO model dependent determination that used
${\cal{B}}(D_S\to\phi\pi^+)=(4.4\pm 0.5)$\%, which is an average
between the PDG value \cite{PDG} and a recent BaBar measurement
\cite{phipi_babar}. In this case $\rho$=0.25. Taking into account
the semileptonic branching ratio (10.5\%), the fraction of high
momentum leptons above the minimum lepton momentum cut (1/3), and
the lepton efficiency (0.8), we estimate that an error of $\pm$4\%
on $f_S$ can be achieved with 30 fb$^{-1}$ of data. (We find that
the fractional error in $f_S$ is about twice the fractional error
in $\rho$.) Interestingly, a new preliminary value of
${\cal{B}}(D_S\to\phi\pi^+)=(3.5\pm 0.4)$\% based on CLEO-c data
\cite{FPCP} raises $f_S$ to 21\%, changes $\rho$ to 0.29, and
consequently improves the sensitivity to about $\pm$3\%. Thus,
this method can lead to a model independent determination of $f_S$
once the C states of the $B^0$ decays are experimentally
determined more precisely.

We next discuss another method for finding the $B_S$ fraction that
uses $B$ mixing as input, but is not the main ingredient. Rather,
we make use of the fact that $D_S^{\pm}$ mesons are produced in
almost all $B_S$ decays, while they are produced only at the 10\%
level in $B$ decays. We define $B\equiv{\cal{B}}(B\to D_S X)$.
This rate has already been measured in $\Upsilon$(4S) decays
\cite{CLEO-DS,PDG}. One complication is that the $B_S$ often
decays into a $D_S^+D_S^-$ pair, which is ignorable in $B$ decays.
To take into account the possibilities of the $B_S$ decaying into
one or two $D_S$ mesons, we define $S_1\equiv{\cal{B}}(B_S\to D_S
X)$, and $S_2\equiv{\cal{B}}(B_S\to D_S^+D_S^- X)$. Here $S_1$ is
the rate for the $B_S$ to decay into one and only one $D_S$ meson,
and $S_2$ is the rate for the $B_S$ to decay into a $D_S$ meson
pair. The inclusive rate ${\cal{B}}(B_S\to D_S X)=S_1+2S_2$. (Note
that the probability that a $B_S$ does not decay into a $D_S$
meson is given by $1-S_1-S_2$.)

In this analysis there are two sources of $D_S$ production, the
$B_S$ again with fraction $f_S$, and $B$ decays, where we take the
mixture of $B^0$ and $B^+$ events to have the same $D_S$ yields as
on the $\Upsilon(4S)$. Since we do not expect the charge of the
$B$ to effect the $D_S$ rates, this method is insensitive to their
relative contribution. Thus, the $B$ fraction is taken as
(1-$f_S$).

Consider now the production of multiple $D_S$ candidates in single
$\Upsilon(5S)$ events, taking into account mixing. $\Upsilon(5S)$
decays can produce events with 4 $D_S$ mesons, when both $B_S$
mesons decay into $D_S^+D_S^- X$. We denote as $N^{+-+-}$ the
observed number of such 4 $D_S$ events and note that this rate is
not effected by mixing. $N^{\pm\pm\mp}$ refers to events with 3
observed $D_S$, whose charge sum is either +1 or -1, and whose
rate is also not effected by mixing. $N^{+-}$ denotes finding an
event with oppositely charge $D_S^+D_S^-$ mesons, whose rate is
changed by both $B_S$ and $B$ mixing, so we introduce the
parameters $f^S_{mix}$ and $f_{mix}$, where $f^S_{mix}$= 1/2 and
$f_{mix}$ equals average mixing rate over the C even, odd,
incoherent mixtures and charged $B$ decays defined above.
$N^{\pm\pm}$ denotes a $D_S^+D_S^+$ or $D_S^-D_S^-$ pair, while
$N^{\pm}$ indicates the detection of a single $D_S^{\pm}$ meson.
Here, the single rate is inclusive of all double, triple and
quadruple rates, etc.. The resulting equations relating the
observed numbers to the branching ratios and $f_S$ are

\begin{eqnarray}
N^{+-+-}/\epsilon^4N_{5S}&=&S_2^2f_S\\
N^{\pm\pm\mp}/\epsilon^3N_{5S}&=&\left(2S_1S_2+4S_2^2\right)f_S\label{eq:triple}\\
N^{\pm\mp}/\epsilon^2N_{5S}&=&\left[(1-f^S_{mix})S_1^2+2(1-f^S_{mix})S_1S_2\right.~~~~~\label{eq:unlike}\\\nonumber
&&\left.+2S_2(1-S_1-S_2)+4S_2^2\right]f_S\\\nonumber
&&+(1-f_S)(1-f_{mix})B^2\\
N^{\pm\pm}/\epsilon^2N_{5S}&=&\left[f^S_{mix}(S_1^2+2S_1S_2)+2S_2^2)\right]f_S~~~~\label{eq:like}\\\nonumber
&&+(1-f_S)B^2f_{mix}\\
N^{\pm}/\epsilon
N_{5S}&=&2(S_1+2S_2)f_S+2(1-f_S)B~,\label{eq:single}
\end{eqnarray}
where $\epsilon$ indicates the detection efficiency of a single
$D_S$; it is the sum of the branching ratio times efficiency for
each decay mode that is used. We ignore two small corrections;
first of all we take the rate for a single $B$ meson to produce a
$D_S^+D_S^-$ pair to be negligibly small, and secondly we don't
account properly for the charge correlations resulting from the
small production of ``wrong-sign" $D_S$ production from $B_S$
decays that can occur via a $b\to u$ transitions. The latter
effects only equations containing $f^S_{mix}$ \cite{fmix}.

We expect that $\epsilon$ could be made as large as 10\% by
including many modes. We use $S_1=0.8$ and $S_2=0.1$, which allows
estimates of the various rates. We do not expect to be able to
observe a significant quadruple $D_S$ rate. On, the other hand, we
estimate that in a 50 fb$^{-1}$ data sample, there would be
$\sim$500 observable triple $D_S$ events allowing the use of
Eq.~\ref{eq:triple}. Thus, we would have four equations, in this
case Eqs.~\ref{eq:triple}-\ref{eq:single}, relating our three
unknowns: $f_S$, $S_1$ and $S_2$. The best values for the unknowns
can be obtained by using a constrained fit to find the solution.
Note that Eq.~\ref{eq:unlike} can be simplified as
\begin{eqnarray}
&&N^{\pm\mp}/\epsilon^2N_{5S}=\left[(1-f^S_{mix})S_1^2-2f^S_{mix}S_1S_2\right.\label{eq:double1}\\\nonumber
&&\left.+2S_2+2S_2^2\right]f_S+(1-f_S)(1-f_{mix})B^2~.
\end{eqnarray}

Should not enough triple $D_S$ events be found in the data sample,
Eqs.~\ref{eq:unlike}-\ref{eq:single} could be solved for the three
unknowns. Another possibility, if precise measurements on the $B^0$
C parity contributions are not available, is to add
Eqs.~\ref{eq:unlike} and \ref{eq:like}, since the resulting equation
is not a function of either $B^0$ or $B_S$ mixing. In this case,
however, measurement of the triple $D_S$ rate is necessary.

In conclusion, knowledge of the number of $B_S$ mesons is essential
for all branching fraction determinations at the $\Upsilon(5S)$. We
present a model independent method of determining the fraction of
$B_S$ mesons at the $\Upsilon(5S)$, and hence
${\cal{B}}\left(\Upsilon(5S) \to
B_S^{(*)}\overline{B}_S^{(*)}\right)$, using the complete mixing of
the $B_S$ via dileptons. The amount of luminosity required to make
an accurate measurement of the $B_S$ fraction will depend on the
actual compositions of the $B^0$ final states, but it is likely to
require several tens of fb$^{-1}$. We also suggest another technique
using the counting of single, double and triple $D_S$ mesons in
$\Upsilon(5S)$ events \cite{otherway}.


This work was supported by the National Science Foundation under
grant \#0553004. We thank Remi Louvot for useful comments.


\begin{thebibliography}{99}
\bibitem{Schneider} Similar techniques have been used at LEP and
Tevatron experiments to help improve the accuracy of the
determination of $b$ hadron fractions. See O. Schneider,
``$B^0\overline{B}^0$ Mixing," in ref.~\cite{PDG}, and references
contained therein.
\bibitem{special}
For better accuracy, final states with a charged and neutral $B$, such as
$B^0B^-\pi^+$ should be put in a separate sum where the coefficient is
${\cal{B}}_{d-sl}{\cal{B}}_{u-sl}$ instead of ${\cal{B}}_{d-sl}^2$. These
states are expected to consitute a small fraction of the decays, so we
have not done this here.
\bibitem{fdequalfc}
In principle, the ratio of $B^-$ to $B^0$ production can be
independently determined by measuring exclusive decays and using
branching ratios measured at the $\Upsilon$(4S). A change in this
ratio from unity will affect the determination of $f_S$ using mixing
but not using inclusive $D_S$ decays.
\bibitem{fmix}
In addition, to take precisely into account any ``wrong-sign" $B$
decays, e. g. $B^-\to D_s^+ X$, we advocate measuring $f_{mix}$ on
the $\Upsilon(4S)$ using the ratio of like-sign to opposite-sign
$D_S$ events.
\bibitem{otherway}

It may also be possible to determine $\overline{B}^0$ and $B^-$
production cross-sections by reconstructing exclusive decays. All
possible modes $B^{(*)}\overline{B}^{(*)}$,
$B^{(*)}\overline{B}^{(*)}\pi$, and $B\overline{B}\pi\pi$ must be
included. The sum of these cross-sections with the $B_S$ yield found
using the techniques described above should equal the total
$\Upsilon$(5S) cross-section. Any short fall would be indicative of
decays of the $\Upsilon$(5S) to final states not containing $b$ and
$\overline{b}$ quarks.
\end{thebibliography}
\end{document}